\documentclass[12pt]{iopart}
\usepackage{amssymb}
\usepackage[dvips]{graphicx}
\newcommand{\Fcal}{{\cal F}}
\newcommand{\Hcal}{{\cal H}}
\newcommand{\Ocal}{{\cal O}}
\newcommand{\qb}{{\bi{q}}}
\newcommand{\rb}{{\bi{r}}}
\newcommand{\ab}{a_{\bot}}
\newcommand{\ee}{\mathrm{e}}
\newcommand\mean[1]{\ensuremath{\langle#1\rangle}}

\begin{document}
% Journal identifier can be put here if required, e.g.
\jl{1}

\title[Variational theory of a model of crystalline tensionless 
surfaces]{Analytical results for a continuum model of crystalline 
tensionless surfaces.  I. Variational mean field study}

\author{Esteban Moro\dag\footnote[2]{Present address:
Theoretical Physics Department, University of Oxford, 1 Keble Road, OX1 3NP,
UK}\footnote[3]{\tt moro@thphys.ox.ac.uk} 
and Rodolfo Cuerno\dag\footnote[4]{\tt cuerno@math.uc3m.es}}

\address{\dag Departamento de Matem\'aticas and Grupo Interdisciplinar de 
Sistemas Complicados, Universidad Carlos III de Madrid, Avda.\ 
Universidad 30, E-28911 Legan\'es, Spain}

\begin{abstract}
We study analytically the equilibrium and near-equilibrium properties of a 
model of surfaces relaxing via linear surface diffusion and subject to a 
lattice potential. We employ the variational mean field formalism introduced
by Saito for the study of the sine Gordon model. In equilibrium, 
our variational theory predicts a first order roughening transition
between a flat low temperature phase and a rough high temperature phase 
with the properties of the linear molecular beam epitaxy equation.
The study of a Gaussian approximation to the Langevin dynamics of the system
indicates that the surface shows hysteresis when we continuously tune 
temperature. Out of equilibrium, this Langevin dynamics approach shows 
that the surface mobility can have different behaviours as a function
of a driving flux. Some considerations are made regarding 
different dimensionalities and underlying lattices, and connections
are drawn to related models or different approaches to the same model we study.
\end{abstract}


\submitted

%\maketitle

\section{Introduction} 
\label{intro}

During the last decade, there have been great theoretical and experimental 
efforts to understand surface growth. This is due to possible 
applications, $e.g.$, to the production of thin films and, from the basic 
point of view, to the interesting examples growing surfaces provide 
of non-equilibrium statistical systems \cite{alb}, in some cases with strong
relation to relevant equilibrium systems \cite{saitobook}.
A very important example is provided by the discrete Gaussian (dG) model, which 
describes the universal features of the equilibrium roughening transition
of many surfaces \cite{weeks}. 
This transition is in the Kosterlitz-Thouless (KT)
class, and thus the model is related to other important models featuring
a similar transition, such as the F model or the Coulomb gas 
\cite{saitobook,weeks}.
The dG model describes a surface minimizing surface area 
(to linear approximation), in which the surface height takes on 
integer values. Relaxing the latter condition leads to the sine-Gordon (sG)
model for a real valued height field subject to surface tension and 
to a (lattice) potential favouring integer values of the field. 
The sG model is amenable to approximate analytic 
treatments \cite{saito78,nozieres1}
which have allowed to develop a rather complete picture of the 
equilibrium roughening transition, and of the surface near-equilibrium 
properties as determined by Langevin dynamics \cite{lang1} or kinetic 
Monte Carlo simulations \cite{lang2}.

There exist surface growth contexts, such as growth by molecular beam epitaxy
(MBE), in which the most relevant relaxation mechanism taking place at the 
surface is surface diffusion \cite{alb}. 
This, in turn, can be modelled as a surface
minimizing curvature, instead of surface area. In reference 
\cite{ourprl}, the following stochastic equation was proposed to study the 
interplay of this mechanism with a lattice potential favouring integer height 
values, similarly to the sG model
\begin{equation}\label{ourmodel}
\frac{\partial h}{\partial t} = F - \kappa \Delta^2 h(\rb) - 
\frac{2\pi V_0}{\ab} \sin \left(\frac{2\pi h(\rb)}{\ab}\right) + \eta(\rb,t).
\end{equation}
In (\ref{ourmodel}), $h(\rb,t)$ is the surface height above (a two-dimensional)
substrate position $\rb$ at time $t$; $\Delta$ is the two dimensional
Laplacian and $\kappa$, $V_0$, and $a_{\bot}$ are positive constants. 
$F$ is also a constant representing, $e.g.$, a driving flux of particles 
inducing the system to grow. $\eta$ is a Gaussian white noise with zero 
average and correlations $\langle\eta(\rb,t)\eta(\rb',t')\rangle = 
2 T \delta(\rb-\rb') \delta(t-t')$, with $T$ being temperature
(we consider a unit Boltzmann's constant, $k_B \equiv 1$). In equilibrium
($F = 0$), this equation governs the thermal fluctuations of a 
surface described by (the continuum limit of) the Hamiltonian
\begin{equation}\label{LrHamcont} \fl
\Hcal = \frac{\kappa}{2} \sum_{i} \left\{\left[\sum_{\delta}
(h_i-h_{i+\delta})\right]^2 + V_0\left[1-\cos\left(\frac{2\pi h_i}{\ab}
\right)\right]\right\},\qquad h_i \in {\mathbb R} .
\end{equation}
where $i+\delta$ denotes a nearest neighbour to site $i$.
In \cite{ourprl}, numerical simulations of (\ref{ourmodel}) showed an
equilibrium roughening transition, similar to that in the sG model; namely, 
for temperatures below a critical value $T_R$ the lattice potential is 
relevant and the surface is flat, whereas for temperatures higher than 
$T_R$ the surface is rough. Out of equilibrium ($F \neq 0$), the surface 
mobility (to be defined below) behaves in different ways, depending on 
$F$ and $T$. Although these simulations have been extended
\cite{tesis,jj}, no analytical approach had been made to study
this model. In this paper we take a first step in this direction and
apply to model (\ref{ourmodel})-(\ref{LrHamcont}) a variational 
mean-field approach successfully applied by Saito \cite{saito78} to the 
study of the sG model. In a subsequent paper we will refine this study by means 
of a dynamic renormalization group (RG) analysis of (\ref{ourmodel}).
The present paper is organized as follows. 
In section \ref{br} we briefly review 
the relationship between (\ref{ourmodel})-(\ref{LrHamcont}) and 
related models for which approximate analytical and/or
numerical results are available. In section \ref{seceq} we study the 
equilibrium Hamiltonian (\ref{LrHamcont}) within the variational scheme of 
\cite{saito78}. 
Section \ref{secdyn} is devoted to the approximate study of the Langevin 
dynamics (\ref{ourmodel}) within a Gaussian approximation for the probability 
distribution of the height. A discussion of the results obtained and our 
conclusions are found in section \ref{concl}. 
Some computational details on the solution of self-consistent
equations relevant to section \ref{seceq} can be found in appendix
\ref{appena}, while appendix \ref{appenb} discusses how the results are 
modified when considering the model on a triangular lattice (as opposed to 
the square lattice studied in the rest of the paper), and appendix
\ref{appenc} contains a discussion on results for substrate dimensions 
different from two.

\section{Background} \label{br}

Crystal surfaces are often described within the solid-on-solid (SOS) 
approximation, in which the surface is characterized by a 
two-dimensional lattice height variable $h_i$ with $h_i/\ab \in {\mathbb Z} $ 
and $i$ being the square lattice position on a $L \times L$ dimensional
substrate. Perhaps the simplest model 
is the discrete Gaussian model (dG) mentioned in the introduction, 
whose Hamiltonian is (we take a unit lattice constant)
\begin{equation}\label{HamDG}
\Hcal_{\rm dG} = \frac{\nu}{2} \sum_{i,\delta} (h_i-h_{i+\delta})^2,\qquad 
h_i /\ab \in {\mathbb Z}
\end{equation}
where $\nu$ is a positive constant. 
Due to the difficulty to handle analytically the discrete sums
in (\ref{HamDG}), a continuum approximation is adopted introducing 
a potential that preserves the periodic symmetry in (\ref{HamDG}) and
favours integer height values, leading to the sine-Gordon (sG) model 
(for reviews, see \cite{saitobook,weeks})
\begin{equation}\label{GenHamcont}
\Hcal_{\rm sG} = \frac{\nu}{2} \sum_{i,\delta} (h_i-h_{i+\delta})^2
+ \sum_{i} V_0\left[1-\cos\left(\frac{2\pi h_i}{\ab}\right)
\right]\qquad h_i \in {\mathbb R} .
\end{equation}
Both models undergo a KT-type roughening transition at a finite 
temperature $T^{\rm sG}_R$ between a flat and a rough phase. 
In the flat phase the roughness $w^2 = (1/L^2) \sum_{i} (h_i-\bar{h})^2$ 
[where $\bar{h}= (1/L^2) \langle\sum_{i} h_i\rangle$] is finite and 
$L$-independent, while 
in the rough phase it diverges with the system size as $w^2 \sim \ln L$,
$i.e.$, as if we take $V_0 = 0$ in (\ref{GenHamcont}). In the flat phase,
the lattice potential dominates and imposes a finite 
correlation length $\xi$. For $T \geq T^{\rm sG}_R$ the lattice potential 
becomes irrelevant, although it modifies \cite{nozieres1} the value of the 
surface tension $\nu$, and the correlation length diverges. Specifically,
$\xi \sim \exp\{C(T_R^{\rm sG}-T)^{-1/2}\}$ for $T \to T_R^{{\rm sG}-}$.

As mentioned in the introduction, our aim is to study surfaces in which 
minimization of surface area is replaced by minimization of surface
curvature, and we will thus replace model (\ref{GenHamcont}) by the 
Hamiltonian (\ref{LrHamcont}) proposed in \cite{ourprl}. Interestingly, in the 
context of two-dimensional melting, Nelson \cite{nelson} proposed 
(on the triangular lattice) the so called Laplacian roughening model 
\begin{equation}\label{LrHam}
\Hcal_{\rm LR} = \frac{\kappa}{2} \sum_{i} \left[\sum_{\delta} 
(h_i-h_{i+\delta})\right]^2,\qquad h_i /\ab \in {\mathbb Z} .
\end{equation}
In equilibrium, numerical simulations and RG studies (see references 
in \cite{strandburg}) indicate that (\ref{LrHam}) displays two 
phase transitions, both in the KT universality class, with an hexatic
phase between the two transition temperatures. However, the conclusions
of analytical and numerical work by other authors (see a review in 
\cite{kleinert}) seem to be that model (\ref{LrHam}) features only one
first order phase transition. Note model (\ref{LrHamcont}) is a natural
continuum approximation of (\ref{LrHam}) in the same spirit as the sG model
is an approximation of the dG model. The numerical study \cite{ourprl}
of Langevin dynamics (\ref{ourmodel}) for (\ref{LrHamcont}) found 
an equilibrium continuous roughening transition between a flat phase 
and a rough phase in which $w^2 \sim L^2$, the same behaviour of the 
so called linear MBE equation [that obtained by setting $V_0 \equiv 0$ in 
(\ref{ourmodel})]. Nevertheless, since the lattice potential modifies
the value of the (in principle zero) surface tension,
the long distance behaviour of the high temperature phase of model 
(\ref{LrHamcont}) was expected to be the same as that in the sG model. 
In what follows we apply Saito's variational treatment to 
(\ref{LrHamcont}) and (\ref{ourmodel}). In the case of the sG model,
such mean field study \cite{saito78} allows to obtain the exact value of the 
roughening temperature, and a rather approximate estimation of the divergence
of the correlation length near $T^{\rm sG}_R$. Thus we expect to obtain 
relevant information from this mean-field scheme. 

\section{Variational mean-field method: equilibrium problem}
\label{seceq}
Following Saito \cite{saito78}, our main assumption is that the most 
relevant features of model 
(\ref{LrHamcont}) can be described by a simpler, solvable Hamiltonian:
\begin{equation}\label{H0}
\Hcal_0 = \frac{T}{2} \sum_{\qb} S^{-1}(\qb) h(\qb) h(-\qb)
\end{equation}
where $h(\qb)$ are the Fourier components of the height field
\begin{equation}\label{hfourier}
h(\qb) = \frac{1}{L}\sum_{j} \ee^{\rmi\qb \cdot \rb_j} h_j .
\end{equation}
Here we consider periodic boundary conditions. Thus, $q_x = 2 \pi n_x/L$ 
with $n_x= -(L-1)/2, \ldots, L/2$ and a similar relation holds for $q_y$. 
Equation (\ref{H0}) defines a Gaussian Hamiltonian in which the values of
$S(\qb)$ are $L^2$ free parameters. We will fix them by minimization of 
the variational free energy $\Fcal_V \equiv \Fcal_0 + \mean{\Hcal - \Hcal_0}_0$,
which is known to be an upper bound of the exact free energy $\Fcal$ of 
model (\ref{LrHamcont}) by the Bogoliubov thermodynamic inequality 
\cite{saitobook}
\begin{equation}\label{bog}
\Fcal \leq \Fcal_V \equiv \Fcal_0 + \mean{\Hcal - \Hcal_0}_0
\end{equation}
where $\Fcal_0$ is the free energy of model $\Hcal_0$ and 
$\mean{\cdots}_0$ stands for the average with respect to the Boltzmann 
factor $\ee^{-\Hcal_0/T}$. 
%Here and in what follows we will use a 
%unit Boltzmann's constant ($k_B \equiv 1$).

Using the Hamiltonians (\ref{LrHamcont}) and (\ref{H0}) we obtain 
for the rhs of equation (\ref{bog})
\begin{eqnarray}
\frac{\Fcal_V}{T}&=& -\frac{1}{2}\sum_{\qb} \ln 2\pi S(\qb) + 
\frac{1}{2} \sum_{\qb} [S_0^{-1}(\qb)-S^{-1}(\qb)] S(\qb) \nonumber \\
& & + \frac{L^2 V_0}{T}
\left\{1-\exp\left(-\frac{2\pi^2}{a_{\bot}^2}w^2\right)\right\} \label{fvlrham} 
\end{eqnarray}
where we have defined $S_0 = T/(\kappa \omega(\qb))$ with $\omega(\qb) = 
16 [\sin^2(q_x/2)+\sin^2(q_y/2)]^2$ and 
\begin{equation}\label{roughness}
w^2 = \frac{1}{L^2} \sum_{j} h_j^2 = \frac{1}{L^2}
\sum_{\qb\neq 0} S(\qb)
\end{equation}
(note that model (\ref{LrHamcont}) is symmetric under $h \to -h$ and thus, 
in equilibrium, $\bar{h} \equiv 0$). 
By minimizing $\Fcal_V$ with respect to the parameters $S(\qb)$,
we find they have to verify
\begin{equation}\label{defsqb}
S^{-1}(\qb) = S^{-1}_0(\qb) + 4\pi^2 \frac{V_0}{a_{\bot}^2 T} 
\exp\left(-\frac{2\pi^2}{a_{\bot}^2} w^2\right) .
\end{equation}
We can rewrite (\ref{defsqb}) by noting that the second term on the rhs 
does not depend on $\qb$. Hence
\begin{equation}\label{sqeq}
S(\qb) = \frac{T}{\kappa(\omega(\qb) + \xi^{-4})}
\end{equation}
where $\xi$ is a constant given by the self-consistent relation [note  
$w^2$ depends on $\xi$ through (\ref{roughness}) and (\ref{sqeq})]
\begin{equation}\label{xieq}
\kappa \xi^{-4} = \frac{4 \pi^2 V_0}{a_{\bot}^2} \exp 
\left(-\frac{2\pi^2}{a_{\bot}^2} w^2\right).
\end{equation}
Equations (\ref{sqeq}) and (\ref{xieq}) are the solution to the equilibrium 
problem. We observe that the variational (Gaussian) approximation 
of Hamiltonian (\ref{LrHamcont}) has a structure factor $S(\qb)$ similar 
to that of the linear MBE equation. 
The only effect of the potential is to introduce a correlation length $\xi$ 
given self-consistently by equations (\ref{xieq}) and 
(\ref{sqeq}). Among all mathematical solutions of equation (\ref{xieq}), 
the best approximation to model (\ref{LrHamcont}) is given by that value of 
$\xi$ that minimizes the variational free energy $\Fcal_V$, 
which we denote by $\xi_{\rm phys}$. Note all roots of equation (\ref{xieq})
can be easily shown to be stationary points of the function $\Fcal_V(\xi)$. 

\begin{figure}
\begin{center}
\includegraphics[width = 0.666\textwidth,height=0.25\textheight]{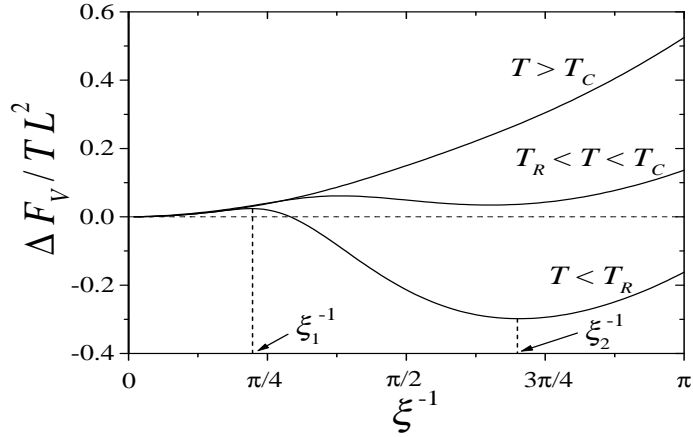} 
\end{center}
\caption{Variational mean free energy difference $\Delta \Fcal_V$ as a
function of the (inverse of the) correlation length for different temperatures.
The values of $\xi_1$ and $\xi_2$ are only displayed for the $T < T_R$ 
case.
The physical value of the correlation length $\xi_{\rm phys}$ is given
by the global minimum of $\Delta \Fcal_V$. For temperatures $T > T_R $
the global minimum is always reached at $\xi^{-1} = 0$.
Parameters used are: $V_0 = \ab = \kappa = 1$.}
\label{fig1}
\end{figure}

\begin{figure}
\begin{center}
\includegraphics[width = 0.666\textwidth,height=0.25\textheight]{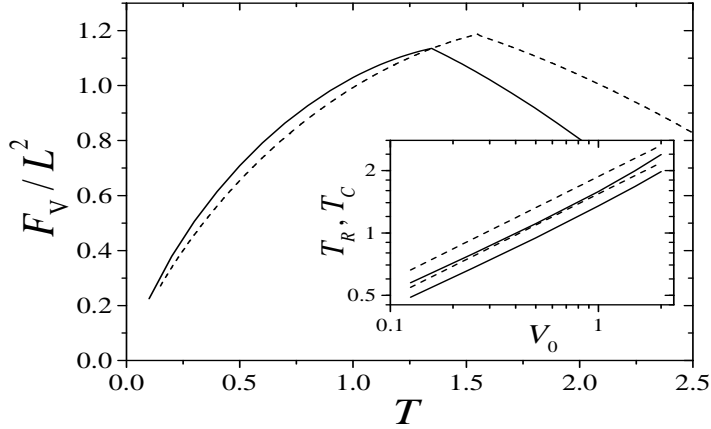}
\end{center}
\caption{Variational free energy $\Fcal_V$ as a function of temperature for
model (\ref{LrHamcont}) using the exact expression (\ref{fvlrham})
(solid line) and our continuum approximation (dashed line).
In both cases, $\Fcal_V$ develops a cusp at $T=T_R$ due to the jump in the
physical value of $\xi$. Inset shows the values of $T_R$
(lower curves) and $T_C$ as functions of $V_0$ within our continuum
approximation (dashed lines) and using the exact discrete expressionis
(solid lines). In both cases, $T_R \propto T_C \sim V_0^{1/2}$.
Parameters used are $\ab = \kappa = 1$.}
\label{fig2}
\end{figure}

In order to proceed analytically, we need to take the continuum limit of
the integrals appearing in (\ref{fvlrham}), (\ref{roughness}). 
In this limit, we make the replacement 
$L^{-2} \sum_{\qb} \rightarrow (2\pi)^{-2}\int 
\rmd \qb$, and we can approximate $\omega(\qb) = \qb^4$,
hence using equation (\ref{sqeq}) we get 
\begin{equation}\label{surrough1}
w^2 \simeq \frac{1}{(2\pi)^2} \int \rmd \qb \frac{T}{\kappa (\omega(\qb) 
+ \xi^{-4})}
= \frac{T \xi^2}{8 \kappa} - \frac{T}{4 \kappa \pi^3} + \Ocal(\xi^{-4}).
\end{equation}
Keeping the dominant term in the above equation (in powers of $\xi$),
and defining $x = 2 \kappa^{1/2} a_{\bot} T^{-1/2}\pi^{-1} \xi^{-1}$ 
equation (\ref{xieq}) becomes
\begin{equation}\label{xieq1}
x^4 = \gamma \ee^{-1/x^2}
\end{equation}
where $\gamma = 64 V_0 a_{\bot}^2 \kappa T^{-2} \pi^{-2}$. 
As shown in appendix \ref{appena}, there are different solutions of equation 
(\ref{xieq1}) depending on $\gamma$ (and therefore on
temperature). Thus, $\xi^{-1} = 0$ is always a solution of (\ref{xieq1}), 
and is the unique solution for $T > T_C = 16 V_0^{1/2} \kappa^{1/2} 
a_{\bot} / \ee \pi$. However, for $T \leq T_C$ there appear two other 
finite solutions $0 < \xi_1^{-1} < \xi_2^{-1}$ of equation (\ref{xieq1}). 
In order to check which of the three roots provides
$\xi_{\rm phys}$ in this temperature range, we compute the free energy 
difference 
\begin{equation}\label{deltaf} \fl
\frac{\Delta \Fcal_V(\xi)}{T L^2} \equiv 
\frac{1}{T L^2}[\Fcal_V(\xi) - \Fcal_V(\xi^{-1} = 0)]
\simeq \frac{\xi^{-2}}{16} - \frac{V_0}{T} \ee^{-T\pi^2 \xi^2 / 
(4 \kappa a_{\bot}^2)} + \Ocal(\xi^{-4}).
\end{equation}
%where const.\ denotes a $\xi$-independent term. 
We plot $\Delta \Fcal_V(\xi)$ in figure \ref{fig1} for different values of $T$.
For $T \leq T_C$, as can be seen in the figure, $\Delta \Fcal_V(\xi)$ has 
indeed a local maximum at $\xi_1^{-1}$ and a local minimum at 
$\xi_2^{-1}$, while for $T > T_C$ both disappear. 
As derived in appendix \ref{appena}, 
for temperatures above $T_R=(\ee^{1/2}/2) \, T_C \simeq 0.82 \, T_C$, 
the variational free energy difference has its global minimum at 
$\xi_{\rm phys}^{-1}=0$.
However, for lower temperatures $T<T_R$, the finite correlation length
$\xi_2$ features a lower value of the variational free energy than the 
infinite correlation length solution, hence $\xi_{\rm phys} = \xi_2$ 
in this temperature range. Summarizing, within the variational approximation
a roughening transition takes place at a temperature
\begin{equation}\label{xieqfin}
T_R = \frac{8}{\pi \ee^{1/2}} \, a_{\bot} \kappa^{1/2} V_0^{1/2} .
\end{equation}
Above $T_R$ the correlation length is infinite and the surface is rough, 
with the same properties as the linear 
MBE model, i.e.\ $S(\qb) \sim \qb^4$ and $w^2 \sim L^2$. Below 
$T_R$ the surface is flat with a finite correlation length equal to $\xi_2$. 
When we approach the roughening temperature from below, 
the correlation length does {\em not} diverge but, rather, tends to a constant 
value (see appendix \ref{appena}) given by
\begin{equation}\label{xieqTR}
\xi(T \to T_R^{-}) = \left(\frac{4\kappa a_{\bot}^2}{T_R \pi^2}\right)^{1/2}
\end{equation}
implying the roughening transition at $T_R$ is of first order. 
Specifically, a cusp develops in the free energy $\Fcal_V$ as a function of 
temperature at $T=T_R$, as depicted in figure \ref{fig2}.

Although the results in this section have been obtained using a certain 
continuum approximation, we have numerically verified all our conclusions
using the exact discrete sums in (\ref{fvlrham}) and (\ref{roughness}). 
The exact variational results for the correlation length and the values of 
$T_C$ and $T_R$ for $L = 1024$ are compared in figure \ref{fig2} to the 
approximate analytical expressions obtained in this section. 
We see that a first order transition indeed takes 
place, although the values of $T_R$ and $T_C$ are modified. However, we
still observe the non-linear dependence of $T_R$ on $V_0$, 
see inset of figure \ref{fig2}. 

\section{Dynamics within the Gaussian approximation}
\label{secdyn}

In this section, we study the near-equilibrium dynamics of model 
(\ref{LrHamcont}) by means of the generalized Langevin growth equation
\begin{equation}\label{Langevin}
\frac{\partial h_i(t)}{\partial t}= F - \frac{\delta \Hcal}{\delta h_i(t)} + 
\eta_i(t)
\end{equation}
where $\eta_i(t)$ is a white noise with correlations
$\langle \eta_i(t) \eta_j(t') \rangle = 2 T \delta_{i,j} \delta(t-t')$ and 
$F$ is the flux of incoming particles in the surface growth picture, or 
a chemical potential difference in a generic context. 
This equation describes not only the 
non-equilibrium statistical dynamics of our model, but also the 
dynamics of the system fluctuations around the equilibrium state (for
$F = 0$). Our approximation \cite{saito78} to the study of equation 
(\ref{ourmodel}) will be to assume a Gaussian time-dependent probability 
distribution for the height field. Thus, we only have to calculate the first 
two moments of the probability distribution, namely the mean 
height\footnote{We make the homogeneity assumption that 
$\langle h_i(t) \rangle$ is independent of substrate position, see 
\cite{saito78}.} $\bar h = \langle h_i(t) \rangle$ and the second moment 
$\langle h(\qb,t) h(-\qb,t)\rangle = S(\qb,t)$. 
Using equations (\ref{LrHamcont}) and (\ref{Langevin}) (see a detailed 
account in \cite{prevar99}) we find 
\begin{eqnarray}\label{meangauss1} \fl
&\frac{\rmd \bar{h}}{\rmd t}& =F - \frac{2\pi V_0}{\ab} 
\left\langle \sin\left(\frac{2\pi h_i}{\ab}\right) \right\rangle \\
\label{meangauss2} \fl
& \frac{\rmd S(\qb,t)}{\rmd t}& =-2 T S(\qb,t) \left[S_0^{-1}(\qb) - 
S^{-1}(\qb,t) +\frac{4\pi^2 V_0}{a_{\bot}^2 T}
\left\langle \cos\left(\frac{2\pi h_i}{\ab}\right)
\right\rangle \right] \\
&  & = -4 T S(\qb,t) \, \frac{\delta \Fcal_V}{\delta S(\qb,t)} \nonumber
\end{eqnarray}
where, within our Gaussian approximation,
\begin{eqnarray}\label{meancossin}
\left\langle \sin\left(\frac{2\pi h}{\ab}\right)\right\rangle & = &
\ee^{-2\pi^2 w^2(t)/a_{\bot}^2}\:\sin\left(\frac{2\pi \bar{h}}{\ab}\right) \\
\left\langle \cos\left(\frac{2\pi h}{\ab}\right)\right\rangle & = &
\ee^{-2\pi^2 w^2(t)/a_{\bot}^2}\:\cos\left(\frac{2\pi \bar{h}}{\ab}\right) 
\end{eqnarray}
with $w^2(t)$ being the time dependent surface roughnes.
%$w^2(t) \equiv \frac{1}{L^2} \sum_{i} (h_i-\bar{h})^2 = 
%\frac{1}{L^2} \sum_{\qb\neq 0} S(\qb,t)$.
In all cases, we will study the set of coupled differential equations
(\ref{meangauss1}) and (\ref{meangauss2}) subject to the initial condition
$h_i(t=0)=0$ for all substrate positions $i$.

\subsection{Equilibrium}
In equilibrium, i.e.\ for $F = 0$, the solution of equation 
(\ref{meangauss1}) is $\bar{h} = 0$ [note (\ref{meancossin})] 
and the solution of (\ref{meangauss2}) is 
the same as that of (\ref{defsqb}) and (\ref{xieq}) obtained in the 
previous section. The interest of equation (\ref{meangauss2}) is that 
it allows us to study dynamically how does the system choose the physical 
value of the correlation length, and corroborate the results obtained in the 
previous section from the point of view of Langevin dynamics. 
Thus, we will integrate numerically 
the complete set of $L^2$ discrete equations (\ref{meangauss1}) and 
(\ref{meangauss2}) and perform the following experiment:
starting from a flat surface and $T = 0$, we increase temperature by a certain
(small) amount and wait until the system reaches equilibrium. 
Then, we increase temperature by the same 
amount and repeat the equilibration process. When the temperature is high 
enough (i.e.\ once the system is in the rough phase) we decrease temperature 
by the same amount and repeat the process of equilibration until 
$T =0$ is reached back closing a temperature cycle.

\begin{figure}
\begin{center}
\includegraphics[width = 0.666\textwidth,height=0.25\textheight]{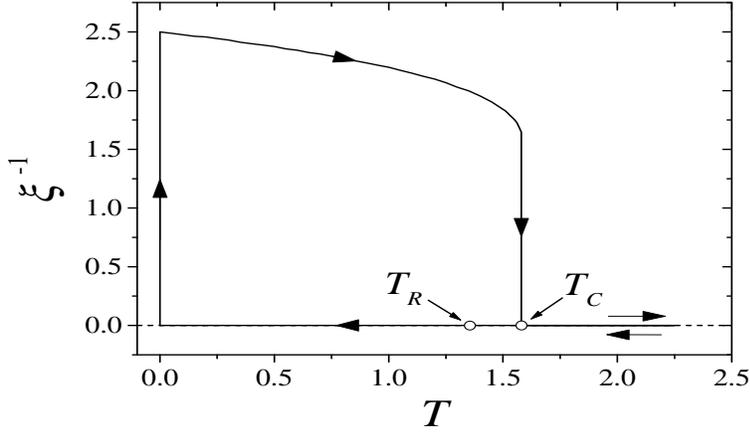}
\end{center}
\caption{(Inverse of the) physical correlation length as a function of 
temperature, as determined from equations (\ref{meangauss2}) and 
(\ref{sqeq}). The arrows indicate the heating and cooling experiment 
explained in the text. Parameters used are $V_0 = a_{\bot} = \kappa = 1$
and $L=1024$.}
\label{fig3}
\end{figure}

We observe that the equilibrium first order transition found
in the previous section indeed induces hysteresis in the system correlation
length (see figure \ref{fig3}) when the system is heated starting 
from $T = 0$, in the sense that the roughening transition takes place at the 
{\em higher} temperature $T_C$ and {\em not} at $T_R$. 
The reason is that, for all $T$
up to $T_C$, the system stays in the local $\Fcal_V$ minimum at $\xi_2$,
even though for $T_R < T < T_C$ the free energy already has its global 
minimum at $\xi^{-1}=0$, since there is an energy barrier for the system to 
jump across the local maximum in $\Delta \Fcal_V$. 
Once the local minimum at $\xi_2$ disappears (i.e.\ for $T \geq 
T_C$), the surface is rough and exhibits an infinite correlation length.
Conversely, when the system is cooled down starting at $T > T_C$, 
the system remains in the rough phase until $T = 0$ is reached because 
$\xi^{-1}=0$ is always a free energy minimum.
%Because the equilibrium configurations are only described by the 
%$\xi^{-1}$ parameter, we could describe the dynamics of the system 
%as the deterministic dynamics of the $\xi^{-1}$ degree of freedom 
%subject to the $\Delta \Fcal(\xi)$ potential. Within this picture, 
%what happens is that the system choose between the two minima for 
%$T \leq T_C$ depending on the initial condition. Looking at figure 
%\ref{fig1}, if we start with a flat initial condition (i.e $\xi = 0$) 
%and because of the existence of a maximum between the two minima, 
%the system rolls down in the free energy landscape until $\xi_2$ is 
%reached. As well as the temperature is increased, the system reacomodates to 
%reach the finite correlation length minimum. When $T_R < T < T_C$, 
%the global minimum is reached at $\xi^{-1} = 0$ but the system still 
%chooses the finite solution because there is a free energy barrier between 
%the two minima. Once $T > T_C$ this barrier disappears and $\xi$ rolls 
%down to $\xi^{-1} =0$. During the cooling, we start with $\xi^{-1}=0$ 
%and because this solution is always a minimum the system remains in that 
%solution until $T=0$.   

\subsection{Non-equilibrium}
\begin{figure}
\begin{center}
\includegraphics[width = 0.666\textwidth,height=0.25\textheight]{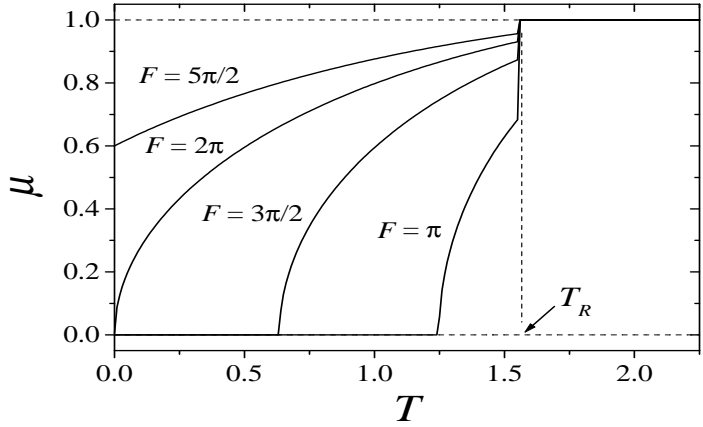}
\end{center}
\caption{Surface mobility as a function of $T$ for different values of the 
driving flux $F$. The values of the mobility are obtained from equation 
(\ref{mobility2}) using equations (\ref{sqeq}) and (\ref{xieq}) with the 
parameter values $V_0 = \kappa = a_{\bot} = 1$.}
\label{fig4}
\end{figure}

In this section we allow $F \neq 0$ in (\ref{meangauss1}) and 
(\ref{meangauss2}), in which case the former no longer has the trivial 
solution ($\bar{h} = 0$). Rather, when the flux $F$ is small 
(quasi-equilibrium condition) we expect the system to feature a structure 
factor $S(\qb,t)$ of the same form as in equilibrium,
all non-equilibrium effects reflecting in the (possibly non-trivial)
behaviour of the average height. Actually, numerical simulations 
\cite{ourprl,tesis} of the 
full non-linear model (\ref{ourmodel}) seem to confirm this expectation. 
For this reason we neglect the feedback effect of the evolution of 
$\bar{h}(t)$ on the structure factor $S(\qb,t)$ and take 
\begin{equation}\label{defsqbnoneq}
S(\qb,t) \simeq \frac{k_B T}{\kappa(\omega(\qb) + \xi^{-4})}
\end{equation}
where $\xi$ is given by the physical equilibrium solution of 
section \ref{seceq}. Within this approximation, $F_c(T) \equiv 
\frac{2 \pi V_0}{a_{\bot}}\exp\{-2\pi^2 w^2/a_{\bot}^2\}$ becomes a constant 
and equation (\ref{meangauss1}) can be written as
\begin{equation}\label{meangauss4}
\frac{\rmd \bar{h}}{\rmd t} = F - F_c \sin \frac{2 \pi \bar{h}}{a_{\bot}} 
\end{equation}
which is simple to integrate analytically 
(exact expressions for the solution can be 
found in \cite{saito78} and \cite{prevar99}). This equation has two 
different solutions depending on the values of $F$. 
If $F \leq F_c$, then $\bar{h}$ tends to a constant value and the surface 
does not grow. If we define the surface mobility $\mu$ as
\begin{equation}
\mu = \frac{1}{F}
\overline{\left\langle \frac{\rmd \bar{h}}{\rmd t}\right\rangle}
\label{mobility2}
\end{equation}
where the overline stands for average over a time larger than $\mu^{-1}$,
then for $F > F_c$ one obtains from the exact solution of (\ref{meangauss4})
a non-zero value for $\mu$:
\begin{equation}\label{mobility}
\mu = \left(1-\frac{F^2_c}{F^2}\right)^{1/2} .
\end{equation}
In figure \ref{fig4} we plot the surface mobility as a function of $T$.
Using the equilibrium solution for $\xi$ described in section \ref{seceq}, 
for temperatures above roughening ($T > T_R$), we have that 
$\xi_{\rm phys}^{-1} = 0$, which implies $F_c = 0$ and $\mu = 1$. Thus,
above roughening the surface shows linear growth with a maximum 
(unit) mobility. In the flat phase ($T< T_R$) the mobility is equal to zero 
(i.e.\ the surface does not move) for a small flux $F < F_c(T)$.
For larger values of the flux ($F > F_c(T)$), the mobility depends 
nonlinearly on $T$ for all temperatures up to $T_R$. 
Due to the jump of the correlation length at $T=T_R$, the mobility also has
a jump at this temperature value. These three behaviours of the surface 
mobility as a function of temperature and driving flux agree with those
obtained \cite{ourprl,tesis} for the full model (\ref{ourmodel}), except
for the discrete jump of $\mu$ at $T=T_R$. 

\section{Discussion and conclusions}
\label{concl}

Summarizing the equilibrium results obtained in the previous sections, 
the variational approximation predicts a first order phase transition for model 
(\ref{LrHamcont}), and the associated hysteresis phenomenon. In particular,
within a Gaussian approximation, 
Langevin dynamics predicts that a rough surface can preserve its 
infinite correlation length when cooled down across the roughening temperature.
Moreover we have found that these results apply both on the square and on the 
triangular lattices (see appendix \ref{appenb}).
Hysteresis behaviour and a first order transition have been reported in
\cite{jk1} and \cite{jk2} for models related with the Laplacian roughening (LR)
model (\ref{LrHam}) on the square lattice and for the LR model on the 
triangular lattice \cite{jt}. However, as mentioned in the introduction,
other authors seem to obtain two KT transitions for the LR model both on 
the square \cite{bruce} and on the triangular \cite{scc} lattices. Note that 
our Langevin dynamics results within the Gaussian approximation yield 
a discrete jump in the surface mobility $\mu$ at $T_R$, which is not found 
in simulations of the full nonlinear model (\ref{ourmodel}). This might 
indicate that the first order character of the transition is in our case
an artifact of the variational approximation. Moreover, this approximation 
(see appendix \ref{appenc}) also predicts a phase transition for model 
(\ref{LrHamcont}) in $d=1$, which is also obtained for the sG model. 
This result points out the limitations of this approximate framework 
for situations in which fluctuations are very relevant for the system 
behaviour (as in the $d=1$ case). Since model (\ref{LrHamcont}) features 
intrinsically strong fluctuations (as does $e.g.$ the linear MBE equation
\cite{alb}), it is desirable to go beyond our present mean field approach
to this model. 
%One relevant possibility is that model (\ref{LrHamcont}) is {\em not} the 
%appropriate real-valued version of the LR model. 
We can take two steps in this direction. 
One (numerical) is to perform extended simulations of both the LR model 
and model (\ref{LrHamcont}) [or, equivalently, its equilibrium Langevin 
dynamics (\ref{ourmodel})]. The results
\cite{jj} seem to indicate that in {\em both} cases there is only one 
continuum transition, though with strong size dependence in the LR case for 
sizes up to moderate (but not large). The other (analytical) improvement
is to perform a dynamic RG analysis of (\ref{ourmodel}) along the lines of 
\cite{nozieres1} for the sG model. This study is particularly important 
bearing in mind that the lattice potential is expected to contribute 
a surface tension, absent in equation (\ref{ourmodel}), which should then
dominate the scaling behaviour as compared with surface diffusion \cite{alb}.
This phenomenon is clearly beyond our mean-field approach, which neglects 
parameter renormalization. Moreover, comparing with the sG case, in the latter
the variational mean field \cite{saito78} and the perturbative
(in powers of $V_0$) RG \cite{nozieres1} approaches
yield the {\em same} (exact) roughening temperature $T_R^{\rm sG} = 
2 \nu a_{\bot}^2/ \pi$, which is independent of $V_0$.
However, in our case $T_R$ does depend on the lattice potential 
(and on $\kappa$) as a fractional power. 
Thus, we do not expect a perturbative RG treatment
to quantitatively agree with the expression for $T_R$ derived here. This 
will be the subject of a forthcoming publication.

\ack

The authors are pleased to thank Angel S\'anchez for discussions, encouragement
and a detailed reading of the manuscript. This work has been partially 
supported by DGES grant No.\ PB96-0119.

\section{Appendix}
\subsection{Solution of the self-consistent equations}
\label{appena}
\begin{figure}
\begin{center}
\includegraphics[width = 0.666\textwidth,height=0.25\textheight]{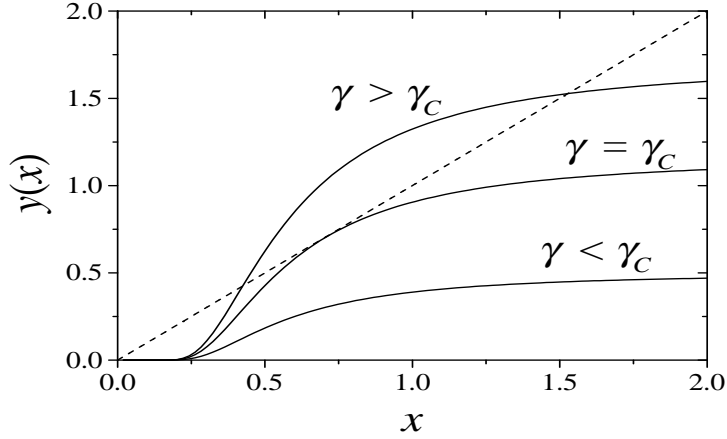}
\end{center}
\caption{Graphical representation of equation (\ref{eqtipo1}). The dashed
line is the $y=x$ function, while the solid lines show $y= \gamma^{1/4}
\ee^{-1/4 x^2}$ for different values of $\gamma$.}
\label{fig5}
\end{figure}

In this Appendix we calculate the self-consistent solution of equations 
(\ref{sqeq}) and (\ref{xieq}) for the equilibrium correlation length 
of the variational approximation (\ref{H0}) to model (\ref{LrHamcont}).
By defining $x = 2 \kappa^{1/2} a_{\bot} T^{-1/2}\pi^{-1} \xi^{-1}$ and
$\gamma = 64 V_0 a_{\bot}^2 \kappa T^{-2} \pi^{-2}$, equations
(\ref{sqeq}) and (\ref{xieq}) become, within the continuum approximation
made in section \ref{seceq} 
\begin{equation}\label{eqtipo}
x^4 = \gamma \ee^{-1/x^2}.
\end{equation}
It is obvious that equation (\ref{eqtipo}) always has the solution $x = 0$, 
and that for some values of $\gamma$ it may also have non-zero solutions. 
Our first aim is to determine the critical value of $\gamma$ for which 
$x = 0$ is the unique solution.
To this end, we rewrite the equation in the following way
\begin{equation}\label{eqtipo1}
x = \gamma^{1/4} \ee^{-1/4 x^2}.
\end{equation}
Now the solutions are the intersections of the function 
$y = f(x) = \gamma^{1/4} \ee^{-1/4 x^2}$ with the straight line $y = x$. 
As we can see in figure \ref{fig5}, 
for $\gamma > \gamma_C$
there are three solutions of equation (\ref{eqtipo1}), two solutions for 
$\gamma = \gamma_C$ and only the trivial solution $x = 0$ for 
$\gamma < \gamma_C$. The value of $\gamma_C$ can be calculated using that 
for $\gamma = \gamma_C$ the unique solution $x = x_s \neq 0$ verifies
(\ref{eqtipo1}) and also the equation
\begin{equation}\label{eqtipo2}
1 = \gamma_C^{1/4}\frac{1}{2 x_s^3} \ee^{-1/4 x_s^2}
\end{equation}
obtained by requiring that the slopes of $y = x$ and $y = f(x)$ be equal 
at $x=x_s$. With these two equations it is easy to obtain 
$x_s = 2^{-1/2}$ and $\gamma_C = \ee^2 / 4$. 
Using the definition of $\gamma$, the temperature for which $\xi^{-1} = 0$ 
is the only solution of equation (\ref{xieq}) is then given by
\begin{equation}\label{Tc}
T_C = \frac{16 V_0^{1/2} \kappa^{1/2} a_{\bot}}{\ee \pi} .
\end{equation}
Now, for $T < T_C$ we have to determine which of the three solutions
of (\ref{eqtipo}) provides the physical correlation length. 
Since $\xi^{-1}=0$ is the unique solution for high temperatures, we 
take as a reference value $\Fcal_V(\xi^{-1}=0)$, and note that
$\Delta \Fcal(\xi) = \Fcal_V(\xi) - \Fcal_V(\xi^{-1}=0)$ is stationary at 
any root of equation (\ref{xieq}). Thus, we will consider as the physical 
solution for the correlation length that root of (\ref{xieq}) for which 
$\Fcal_V$ has an absolute minimum. Starting out with high temperatures, 
the condition $\Delta \Fcal_V(\xi) = 0$
will signal the temperature at (and below) which $\xi^{-1}=0$ ceases 
to be the global minimum of the variational free energy and thus the 
system physical correlation length. Using our previous notation, 
the condition $\Delta \Fcal_V(\xi) = 0$ reads
\begin{equation}\label{eqtipo3}
x^2 = \gamma' \ee^{-1/x^2}
\end{equation}
where $\gamma'=64 \kappa a_{\bot}^2 V_0/(\pi^2 T^2)$. Using
the same argument as above, it is easy to show that for 
$\gamma' < \gamma'_R = \ee$ there are non-zero solutions of (\ref{eqtipo3}). 
This means, using the definition of $\gamma'$, that there is a temperature 
given by
\begin{equation}\label{TR}
T_R = \frac{\ee^{1/2}}{2} T_C,
\end{equation}
such that for $T < T_R$ the global minimum of the free energy is attained
for a correlation length $\xi \neq 0$, whereas for 
$T \geq T_R$ the physical solution is $\xi_{\rm phys}^{-1} = 0$.

\subsection{Triangular lattice}
\label{appenb}

The Laplacian roughening model was initially proposed by Nelson on the 
triangular lattice \cite{nelson}. Thus, it is worth studying how do the 
features of our 
model (\ref{LrHamcont}) change when the substrate geometry is different
from the square lattice considered in the text. Nevertheless, we expect 
that only nonuniversal quantities ---such as the transition temperature and 
the numerical value of correlation length--- depend upon the lattice geometry.
The Laplacian roughening model on the triangular lattice is given by 
\begin{equation}
\Hcal_{\rm LR} = \frac{\kappa}{2} \sum_i \left[\sum_{\delta}
(h_i-h_{i+\delta})\right]^2
\end{equation}
with $i+\delta$ being any of the six nearest neighbours of site $i$. 
For this case \cite{prevar99},
$\omega(\qb) = 16 
\{\sin^2(q_x/2)+\sin^2[(q_x+\sqrt{3}q_y)/4]+\sin^2[(q_x-\sqrt{3}q_y)/4]\}^2$,
where $\qb = (n_x/L) \mathbf{b}_x+(n_y/L) \mathbf{b}_y$, with 
$\mathbf{b}_x = 2 \pi[\mathbf{e}_x-(1/\sqrt{3})\mathbf{e}_y]$ and 
$\mathbf{b}_y = (4\pi/\sqrt{3}) \mathbf{e}_y$, where $n_i=-(L-1)/2,\ldots,L/2$
and $\mathbf{e}_i$ are the standard basis vectors.
In the continuum limit, $S_0(\qb) \simeq 4 T/(9 \kappa \qb^4)$, and we 
recover equation (\ref{xieq}). Taking the continuum
limit (i.e $\frac{1}{L^2}\sum_{\qb} \to \frac{\sqrt{3}}{2} \int_{BZ} 
\frac{\rmd^2 \qb}{(2\pi)^2} \simeq 
\frac{\sqrt{3}}{4\pi}\int_{0}^{(2/\sqrt{3})^{1/2} \pi} q \, \rmd q$, 
where $BZ$ denotes the first Brillouin zone), we get
\begin{equation}\label{roughtri}
w^2 \simeq \frac{T}{2 \pi \kappa (2/\sqrt{3})^2} \int_0^{\pi} 
\frac{q'}{q'^4+ \frac{\xi^{-4}}{(2/\sqrt{3})^2}} \rmd q' .
\end{equation}
Thus, by defining $T' = T / 3$, and $\xi' = 
3^{1/4} \xi$, we get the same equation 
(\ref{xieq1}) but with redefined constants $T'$ and $\xi'$. One can readily 
reproduce all the results obtained in the text, simply by making the 
replacements $T \to T'$ and $\xi \to \xi'$. In conclusion, on the triangular
lattice a first order roughening transition is also obtained, the only effect 
of the geometry being a shift in the value of the roughening temperature
$T_R^{\rm triang.}= T_R^{\rm square}/3$.

\subsection{Substrate dimensions $d \neq 2$}
\label{appenc}

In this appendix we discuss the possibility to find a roughening transition in
equilibrium when model (\ref{LrHamcont}) is defined on a substrate of 
generic dimension $d$. In that case, equation
(\ref{xieq}) is still valid, but with
\begin{equation}\label{roughddim}
w^2(\xi) \simeq \int \frac{\rmd^d \qb}{(2 \pi)^d} \frac{T}{\kappa ( \omega(\qb) + \xi^{-4})},
\end{equation}
within the continuum limit. For substrate dimension $d > 4$, the integral
(\ref{roughddim}) is finite for $\xi^{-1}=0$, namely
$w^2(\xi^{-1} = 0) = K_d \pi^{d-4} T/ [\kappa (d-4)]$
(where $K_d$ is the $d$-dimensional angular integral 
$K_d = \int \rmd^{d-1}\Omega/(2\pi)^d = 2 \pi^{d/2}/[(2 \pi)^d \Gamma(d/2)]$).
Thus, $\xi^{-1} = 0$ is no longer a solution of equation (\ref{xieq}).
Therefore the system has no rough solution and is in the flat phase for all 
temperatures. On the other hand, for $d < 4$, the integral
above may be approximated by
\begin{equation}\label{approxwddim}
w^2(\xi) \lesssim \frac{T}{\kappa} 
\frac{\pi K_d}{4 \sin(\pi d / 4)}\ \xi^{4-d}.
\end{equation}
In this case, $\xi^{-1} = 0$ is always solution of equation (\ref{xieq}), 
there being two additional finite solutions when $T < T_C^d$. 
The value of $T_C^d$ can be calculated using the same argument as in 
$d = 2$ and is
\begin{equation}\label{tcd}
T_C^{d<4} = \frac{8 \kappa a_{\bot}^2 \sin(d\pi/4)}{(4-d) K_d\: \ee\: \pi^3}
\left(\frac{4 \pi^2 V_0}{\kappa a_{\bot}^2}\right)^{\frac{4-d}{4}}.
\end{equation}
In order to know which solution of equation (\ref{xieq}) minimizes the 
variational free energy, we calculate $\Delta \Fcal_V$, which now reads
\begin{equation}\label{deltafd}
\frac{\Delta \Fcal_V(\xi)}{T L^d} \simeq \frac{4-d}{d} 
\frac{\pi K_d}{8 \sin(\pi d/4)}\: \xi^{-d}
-\frac{V_0}{T} \ee^{-2\pi^2 w^2 / a_{\bot}^2}.
\end{equation}
In this case, the local minimum $\xi^{-1}_2 \neq 0$ is also the global 
minimum and the physical solution for temperatures below the roughening
temperature ($T<T_R^{d<4}$), which is given by
\begin{equation}\label{trd}
T_R^{d<4} = \frac{d}{4} \ee^{\frac{4-d}{4}} \, T_C^{d<4}.
\end{equation}
For temperatures above roughening ($T \geq T_R^{d<4}$), $\xi^{-1} = 0$ 
provides the global free energy minimum. Thus, for $d < 4$
there is a first order roughening transition at $T_R^{d<4}$.
Note this includes $d=1$, which might seem conflictive since in this case
model (\ref{LrHamcont}) is expected to be in the rough phase for all values 
of $T$ \cite{nelson}: In $d=1$ thermal fluctuations are expected to destroy 
the ordered flat phase for any temperature value.
Our result can be understood by noting that $\Fcal_V$ is {\em not}
a true free energy, in the sense that it is not the free energy of any 
model\footnote{The free energy of model $\Hcal_0$ is $\Fcal_0$.}, but rather 
an upper bound for the free energy of model (\ref{LrHamcont}). 
Actually, one obtains exactly the same result in the 
variational study of the sG model in $d=1$ \cite{saito78}. Note that in this
reference the analysis of the $d=1$ case is incomplete, with the incorrect 
conclusion that the variational theory predicts no phase transition when $d=1$.
The complete expression for $\Delta \Fcal_V(\xi)$ analogous to (\ref{deltafd})
indeed shows that also for the sG model in $d=1$ the variational approximation 
does predict a non-zero temperature below which the physical value of the 
correlation length is finite. 

Finally, for $d=4$ equation (\ref{xieq}) is very similar to that obtained by 
Saito for the sine-Gordon model
\begin{equation}\label{xieqd}
\kappa \xi^{-4} = \frac{4\pi^2 V_0}{a_{\bot}^2}\left(1+\pi^4 \xi^4
\right)^{-\frac{T}{16 \kappa a_{\bot}^2}}.
\end{equation}
Following Saito's analysis for the sine-Gordon model \cite{saitobook}, we 
readily obtain that for $d=4$ our model has a Kosterlitz-Thouless transition 
when $T = T_R^{d=4} \equiv 16\kappa a_{\bot}^2$. The
correlation length now diverges as $\xi \sim \exp\{-A/(T-T_R^{d=4})\}$ 
when $T \to T_R^{d=4,-}$ ($A$ is a $T$ independent constant).

In summary, within the variational approach, 
our model displays a first order transition for $d < 4$ 
between a flat phase and a rough phase with the properties of the 
linear MBE equation. For the marginal dimension $d = 4$ this transition 
becomes of the Kosterlitz-Thouless type, whereas for $d > 4$ the surface 
is in the flat phase for all temperature values.

\end{document}